\documentclass[preprint,showpacs,superscriptaddress,nofootinbib]{revtex4}
\usepackage[dvips]{graphicx}
\usepackage{amsmath,amssymb,cancel}

\newcommand{\eV}{{\text{eV}}}
\newcommand{\keV}{{\text{keV}}}

\newcommand{\MeV}{{\text{MeV}}}
\newcommand{\GeV}{{\text{GeV}}}
\newcommand{\TeV}{{\text{TeV}}}

\newcommand{\tr}{{\text{tr}}}
\newcommand{\HTM}{{\text{HTM}}}
\newcommand{\pell}{{\ell^\prime}}
\newcommand{\dprime}{{\prime\prime}}
\newcommand{\BL}{{\text{B}-\text{L}}}

\newcommand{\SU}{{\text{SU}}}
\newcommand{\qqHppHmm}
           {{q\overline{q} \to Z^\ast (\gamma^\ast) \to H^{++}H^{--}}}
\newcommand{\qqHppHm}
           {{q^\prime\overline{q} \to W^{\pm\ast} \to H^{\pm\pm}H^\mp}}

\begin{document}
\preprint{UT-HET 065}

\title{
Dark matter and a suppression mechanism
for neutrino masses\\
in the Higgs triplet model}

\author{Shinya Kanemura}
\email{kanemu@sci.u-toyama.ac.jp}
\affiliation{
Department of Physics,
University of Toyama, Toyama 930-8555, Japan
}
\author{Hiroaki Sugiyama}
\email{hiroaki@fc.ritsumei.ac.jp}
\affiliation{
Department of Physics,
Ritsumeikan University, Kusatsu, Shiga 525-8577, Japan
}


\begin{abstract}
 We extend the Higgs triplet model so as to include dark matter candidates
 and a simple suppression mechanism for the vacuum expectation value ($v_\Delta$)
 of the triplet scalar field.
 The smallness of neutrino masses can be naturally
 explained with the suppressed value of $v_\Delta$
 even when the triplet fields are at the TeV scale. 
 The Higgs sector is extended by introducing $Z_2$-odd scalars 
 (an $\SU (2)_L$ doublet $\eta$ and a real singlet $s_2^0$)
 in addition to a $Z_2$-even complex singlet scalar $s_1^0$
 whose vacuum expectation value violates the lepton number conservation by a unit.
 In our model, $v_\Delta$ is generated by the one-loop diagram
 to which $Z_2$-odd particles contribute.
 The lightest $Z_2$-odd scalar boson
 can be a candidate for the dark matter.
 We briefly discuss a characteristic signal of our model at the LHC\@.
\end{abstract}

%

\maketitle

\section{introduction}
\label{sec:intro}

 Existence of dark matter~(DM) has been established,   
 and its thermal relic abundance has been determined
 by the WMAP experiment~\cite{WMAP, Komatsu:2008hk}.
 If the essence of DM is an elementary particle,
 the weakly interacting massive particle~(WIMP) would be a promising
 candidate. 
 It is desired to have a viable candidate for the dark matter
 in models beyond the standard model~(SM).
 The WIMP dark matter candidate
can be accommodated economically
by introducing only an inert scalar field%
~\cite{Silveira:1985rk, Deshpande:1977rw, i-nplet},
where we use ``inert'' for the $Z_2$-odd property.
 The imposed $Z_2$ parity ensures the stability of the DM candidate.
 Phenomenology in such models have been studied in,
e.g., Refs.~\cite{c-i-singlet, r-i-singlet,
i-doublet, Lundstrom:2008ai,Araki:2011hm, THDM-iSDM, HTM-iSDM}.

 On the other hand,
it has been confirmed by neutrino oscillation measurements
that neutrinos have nonzero but tiny masses
as compared to the electroweak scale%
~\cite{solar, atm, acc, reactor-S, reactor-L}.
 The different flavor structure of neutrinos
from that of quarks and leptons
may indicate that neutrino masses are of Majorana type.
 In order to explain tiny neutrino masses,
many models have been proposed.
 The seesaw mechanism is the simplest way
to explain tiny neutrino masses,
in which right-handed neutrinos are introduced
with large Majorana masses~\cite{seesaw,Mohapatra:1979ia}.
 Another simple model for generating neutrino masses 
is the Higgs Triplet Model~(HTM)~\cite{Mohapatra:1979ia, HTM}.
 However,
these scenarios do not contain dark matter candidate in themselves.

 In a class of models where tiny neutrino masses are generated by
 higher orders of perturbation,
the DM candidate can be naturally contained%
~\cite{KNT, Ma, AKX, Aoki:2011yk, Kanemura:2010bq, Gu:2007ug}. 
 In models in Refs.~\cite{KNT, Ma, AKX, Aoki:2011yk, Kanemura:2010bq},
the Yukawa couplings of neutrinos with the SM Higgs boson
are forbidden at the tree level
by imposing a $Z_2$ parity.
 The same $Z_2$ parity also guarantees
the stability of the lightest $Z_2$-odd particle in the model
which can be the candidate of the DM
as long as it is electrically neutral.

 In this paper, we consider an extension of the HTM
in which by introducing the $Z_2$ parity
$m_\nu$ is generated at the one-loop level
and the DM candidate appears.
 In the HTM, Majorana masses for neutrinos are generated via the Yukawa interaction 
 $h_{\ell\pell} \overline{L_\ell^c}\, i\sigma_2 \Delta L_\pell$ 
 with a nonzero vacuum expectation value~(VEV) of an
 $\SU (2)_L$ triplet scalar field $\Delta$ with the hypercharge of
 $Y=1$.
 The VEV of $\Delta$ is described by
$v_\Delta \sim \sqrt{2} \mu v^2/(2 M_\Delta^2)$,
where $v$ is the VEV of the Higgs doublet field $\Phi$
and $M_\Delta$ is the typical mass scale of the triplet field;
 the dimensionful parameter $\mu$
breaks lepton number conservation
at the trilinear term $\mu\,\Phi^T i\sigma_2 \Delta^\dagger \Phi$
which we refer to as the $\mu$-term.
 As the simplest explanation for the smallness of neutrino masses, 
 the mass of the triplet field is assumed to be much larger than the
 electroweak scale.
  On the other hand a characteristic feature of the HTM is the fact that 
 the structure of the neutrino mass matrix $(m_\nu)_{\ell\pell}$ is
 given by that of the Yukawa matrix,  
 $h_{\ell\pell} \propto (m_\nu)_{\ell\pell}$.  
 The direct information on $(m_\nu)_{\ell\pell}$
would be extracted from the decay 
$H^{\pm\pm} \to \ell^\pm\pell^{\prime\pm}$~\cite{nu-LHC}
if $H^{++}$ is light enough to be produced at collider experiments,
where $H^{++}$ is
the doubly charged component of the triplet field $\Delta$.
 At hadron colliders,
the $H^{\pm\pm}$ can be produced via
$\qqHppHmm$~\cite{HppHmm} and $\qqHppHm$~\cite{HppHm}.
The $H^{\pm\pm}$ searches at the LHC put lower bound on its mass as
$m_{H^{\pm\pm}}^{}\gtrsim 300\,\GeV$~\cite{Hpp-CMS,Hpp-ATLAS},
assuming that the main decay mode is
$H^{\pm\pm} \to \ell^\pm \ell^{\prime\pm}$. 
 Phenomenological analyses
for $H^{\pm\pm}$ in the HTM at the LHC 
have also been performed in Ref.~\cite{Hpp}.
 Triplet scalars can contribute to lepton flavor violation~(LFV)
in decays of charged leptons,
e.g., $\mu \to \bar{e}ee$
and $\tau \to \bar{\ell}\pell\ell^{\dprime}$ at the tree level
and $\ell \to \pell \gamma$ at the one-loop level.
 Relation between these LFV decays and neutrino mass matrix
constrained by oscillation data
was discussed in Refs.~\cite{Chun:2003ej,LFV-HTM}.
 In order to explain the small $v_\Delta$
with such a detectable light $H^{++}$,
the $\mu$ parameter has to be taken to be unnaturally  
 much lower than the electroweak scale.
 Therefore, it would be interesting to extend the HTM
in order to include a natural suppression mechanism
of the $\mu$ parameter (therefore $v_\Delta$)
in addition to the DM candidate.


 In our model,
lepton number conservation is imposed to the Lagrangian
in order to forbid the $\mu$-term in the HTM at the tree level
while the triplet Yukawa term
$h_{\ell\pell}^{} \overline{L_\ell^c}\, i\sigma_2 \Delta L$
exists.
 The VEV of a $Z_2$-even complex singlet scalar $s_1^0$
breaks the lepton number conservation by a unit.
 An $\SU (2)_L$ doublet $\eta$ and a real singlet $s_2^0$
are also introduced as $Z_2$-odd scalars
in order to accommodate the DM candidate.
 Then,
the $\mu$-term is generated at the one-loop level
by the diagram in which the $Z_2$-odd scalars
are in the loop.
 By this mechanism,
the smallness of $v_\Delta \ll v$ is realized,
and the tiny neutrino masses are naturally explained
without assuming the triplet fields to be heavy.
 The Yukawa sector is then the same as the one in the HTM, 
so that its predictions for the LFV processes are not changed.
 See Refs.~\cite{Babu:2001ex,Chun:2003ej}
for some discussions
about two-loop realization of the $\mu$-term%
\footnote{
 The two-loop $\mu$-term in Ref.~\cite{Babu:2001ex}
is given with softly-broken $Z_4$ symmetry,
but the tree level $\mu$-term would be also accepted
as a soft breaking term.
 The two-loop $\mu$-term in Ref.~\cite{Chun:2003ej}
is given with $Z_3$ symmetry which is broken
by a VEV of a scalar $S$,
but the tree level $\Phi^T\, i\sigma_2 \Delta^\dagger \Phi S^\ast$
seems allowed by the $Z_3$.
}.

 This paper is organized as follows.
 In Sec.~\ref{sec:HTM},
we give a quick review for the HTM to define notation.
 In Sec.~\ref{sec:1-loop},
the model for radiatively generating the $\mu$ parameter
with the dark matter candidate
is presented.
 Some phenomenological implications are discussed
in Sec.~\ref{sec:pheno},
and the conclusion is given in Sec.~\ref{sec:concl}.
 The full expressions of the Higgs potential
and mass formulae for scalar bosons in our model
are given in Appendix.

\section{Higgs Triplet Model}
\label{sec:HTM}

 In the HTM,
an $\text{SU}(2)_L$ triplet of complex scalar fields
with hyperchage $Y=1$
is introduced to the SM\@.
 The triplet $\Delta$
can be expressed as
\begin{eqnarray}
\Delta
&=&
 \begin{pmatrix}
  \Delta^+/\sqrt{2} & \Delta^{++}\\
  \Delta^0 & -\Delta^+/\sqrt{2}
 \end{pmatrix} ,
\end{eqnarray}
where $\Delta^0 = (\Delta^0_r + i \Delta^0_i)/\sqrt{2}$.
 The triplet has a new Yukawa interaction term with leptons as
\begin{eqnarray}
{\mathcal L}_{\text{triplet-Yukawa}}
&=&
 h_{\ell\pell}\, \overline{L_\ell^c}\, i\sigma_2\, \Delta\, L_\pell
 + \text{h.c.} ,
\end{eqnarray}
where $h_{\ell\pell}$~($\ell, \pell = e, \mu, \tau$)
are the new Yukawa coupling constants,
$L_\ell$~[$= (\nu_{\ell L}, \ell)^T$] are lepton doublet fields,
a superscript $c$ means the charge conjugation,
and $\sigma_i$~($i = 1\text{-}3$) denote the Pauli matrices.
 Lepton number~($L\#$) of $\Delta$ is
assigned to be $-2$ as a convention
such that the Yukawa term does not break the conservation.
 A vacuum expectation value
$v_\Delta^{}$~[$=\sqrt{2}\,\langle\Delta^0\rangle$]
breaks lepton number conservation by two units.
 The new Yukawa interaction then yields
the Majorana neutrino mass term
$(m_\nu)_{\ell\pell}\,\overline{(\nu_{\ell L}^{})^c}\,\nu_{\pell L}^{}/2$
where $(m_\nu)_{\ell\pell} = \sqrt{2}\, v_\Delta\, h_{\ell\pell}$.

 The scalar potential in the HTM
can be written as
\begin{eqnarray}
V_\HTM
&=&
 -m_\Phi^2\, \Phi^\dagger \Phi
 + m_\Delta^2 \tr(\Delta^\dagger \Delta)
 + \left\{
    \mu\, \Phi^T i\sigma_2 \Delta^\dagger \Phi
    + \text{h.c.}
   \right\}
\nonumber\\
&&{}
 + \lambda_1 (\Phi^\dagger \Phi)^2
 + \lambda_2
   \left[
    \tr(\Delta^\dagger \Delta)
   \right]^2
 + \lambda_3\, \tr[(\Delta^\dagger \Delta)^2]
\nonumber\\
&&{}
 + \lambda_4\, (\Phi^\dagger \Phi) \tr(\Delta^\dagger \Delta)
 + \lambda_5\, \Phi^\dagger \Delta \Delta^\dagger \Phi ,
\end{eqnarray}
where $\Phi=(\phi^+, \phi^0)^T$~[$\phi^0=(\phi^0_r + i\phi^0_i)/\sqrt{2}$\,]
is the Higgs doublet field in the SM\@.
 The $\mu$ parameter can be real
by using rephasing of $\Delta$.
 Because we take $m_\Delta^2 > 0$,
there is no Nambu-Goldstone boson
for spontaneous breaking
of lepton number conservation.
 The small triplet VEV $v_\Delta^{}$ is generated
by an explicit breaking parameter $\mu$
of the lepton number conservation as
\begin{eqnarray}
v_\Delta^{}
&\simeq&
 \frac{ \sqrt{2}\, \mu v^2 }
      { 2m_\Delta^2 + (\lambda_4+\lambda_5) v^2 } ,
\end{eqnarray}
where $v$~($\simeq 246\,\GeV$) is
the doublet VEV defined by $v=\sqrt{2}\,\langle\phi^0\rangle$. 

 In order to obtain small neutrino masses in the HTM,
at least one of $v^2/m_\Delta^2$, $h_{\ell\pell}$, $\mu/v$
should be tiny.
 A small $\mu$ is an attractive option
because $m_\Delta^{}$ can be small ($\lesssim 1\,\TeV$)
so that triplet scalars can be produced at the LHC\@.
 Furthermore,
large $h_{\ell\pell}$ can be taken,
which have direct information on
the flavor structure of $(m_\nu)_{\ell\pell}$.
 There is, however,
no reason why the $\mu$ parameter is tiny in the HTM\@.
 In our model presented below,
the $\mu$ parameter is naturally small
because it arises at the one-loop level.

\section{An extension of the Higgs Triplet Model}
\label{sec:1-loop}

 Since we try to generate the $\mu$-term in the HTM radiatively,
the term must be forbidden at the tree level.
 The simplest way would be
to impose lepton number conservation to the Lagrangian.
 The conservation is assumed to be broken
by the VEV of a new scalar field $s_1^0$
which is singlet under the SM gauge symmetry.
 Notice that $s_1^0$~[$= (s_{1r}^0 + i s_{1i}^0)/\sqrt{2}$] 
is a complex ("charged") field with non-zero lepton number
although it is electrically neutral.
 One might think that
the VEV of $s_1^0$ could be generated
by using soft breaking terms of $L\#$.
 However,
the $\mu$-term is also a soft breaking term.
 Therefore
lepton number must be broken spontaneously in our scenario.
 One may worry about Nambu-Goldstone boson
corresponds to the spontaneous breaking
of the lepton number conservation
(the so-called Majoron, $J^0$).
 However
the Majoron which comes from gauge singlet field
can evade experimental searches (constraints)
because it interacts very weakly with matter fields~\cite{majoron}.
 It is also possible to make it absorbed by a gauge boson
by introducing the $\text{U}(1)_\BL$ gauge symmetry
to the model (See, e.g., Ref.~\cite{BL}).
 In this paper
we just accept the Majoron
without assuming the $\text{U}(1)_\BL$ gauge symmetry
for simplicity.

 If $s_1^0$ has $L\# = -2$,
we can have a dimension-4 operator
$\lambda\, s_1^0\, \Phi^T i\sigma_2 \Delta^\dagger \Phi$.
 This gives a trivial result
$\mu = \lambda \langle s_1^0 \rangle$ at the tree level.
 Although the dim.-4 operator could be forbidden
by some extra global symmetries with extra scalars to break them,
we do not take such a possibility in this paper.
 We just assume the $s_1^0$ has $L\#=-1$.
 Then
the lepton number conserving operator
which results in the $\mu$-term is of dimension-5 as
\begin{eqnarray}
(s_1^0)^2 \Phi^T i\sigma_2 \Delta^\dagger \Phi .
\label{eq:dim5op}
\end{eqnarray}
 We consider below how to obtain the dim.-5 operator
at the loop level by using renormalizable interactions%
\footnote
{
 It will not be difficult to do the same consideration
for cases of higher dimensional operators,
e.g., dim.-6 one with $s_1^0$ of $L\#=-2/3$.
}.
 We restrict ourselves to extend
only the $\text{SU}(3)_c$-singlet scalar sector in the HTM
because it seems a kind of beauty that
the HTM does not extend the fermion sector and colored sector in the SM\@.
 An unbroken $Z_2$ symmetry is introduced
in order to obtain dark matter candidates,
and new scalars
which appear in the loop diagram for the $\mu$-term
are aligned to be $Z_2$-odd particles.
 We emphasize that the unbroken $A_2$ symmetry
is not for a single purpose to introduce dark matter candidates
but utilized also for our radiative mechanism for the $\mu$-term.

\begin{table}[t]
\begin{center}
\begin{tabular}{c|c|c|c||c|c|c}
 {}
 & \ $L$ \
 & \ $\Phi$ \
 & \ $\Delta$ \
 & \ $s_1^0$ \
 & \ $s_2^0$ \
 & \ $\eta$ \
\\\hline\hline
 \ $\text{SU}(2)_L$ \
 & \ {\bf \underline{2}} \
 & \ {\bf \underline{2}} \
 & \ {\bf \underline{3}} \
 & \ {\bf \underline{1}} \
 & \ {\bf \underline{1}} \
 & \ {\bf \underline{2}} \
\\\hline
 \ $\text{U}(1)_Y$ \
 & \ $1/2$ \
 & \ $1/2$ \
 & \ $1$ \
 & \ $0$ \
 & \ $0$ \
 & \ $1/2$ \
\\\hline
 \ $L\#$ \
 & \ $1$ \
 & \ $0$ \
 & \ $-2$ \
 & \ $-1$ \
 & \ $0$ \
 & \ $-1$ \
\\\hline
 \ $Z_2$ \
 & \ $+$ \
 & \ $+$ \
 & \ $+$ \
 & \ $+$ \
 & \ $-$ \
 & \ $-$ \
\end{tabular}
\end{center}
\caption{
 List of particle contents of our one-loop model.
}
\label{tab:1-loop}
\end{table}

 We present the minimal model where
the dim.-5 operator in eq.~(\ref{eq:dim5op})
is generated by a one-loop diagram with dark matter candidates.
 Table~\ref{tab:1-loop} shows the particle contents.
 A real singlet scalar field $s_2^0$
and the second doublet scalar field
$\eta$~[$=(\eta^+, \eta^0)^T, \eta^0=(\eta^0_r+i\eta^0_i)/\sqrt{2}$\,]
are introduced to the HTM in addition to $s_1^0$.
 Lepton numbers of $s_2^0$ and $\eta$ are 0 and $-1$, respectively.
 Then
$\eta^T i\sigma_2 \Delta^\dagger \eta$
conserves lepton number.
 In order to forbid the VEV of $\eta$,
we introduce an unbroken $Z_2$ symmetry
for which $s_2^0$ and $\eta$ are odd.
 Other fields are even under the $Z_2$.

 The Yukawa interactions are the same as those in the HTM\@.
 The Higgs potential is given as
\begin{eqnarray}
V
&=&
 \frac{1}{\,2\,} m_{s_2^0}^2 (s_2^0)^2
 + \left\{
    \mu_\eta^{}\, \eta^T\, i\sigma_2\, \Delta^\dagger\, \eta
    + \text{h.c.}
   \right\}
 + \left\{
    \lambda_{s\Phi\eta}\, s_1^0\, s_2^0\, (\eta^\dagger\, \Phi)
    + \text{h.c.}
   \right\}
 + \cdots .
\end{eqnarray}
 Here
we show only relevant parts
for radiative generation of the $\mu$-term.
 See Appendix for the other terms.
 Vacuum expectation values $v$
and $v_s$~[$= \sqrt{2}\,\langle s_1^0 \rangle$] are given by
\begin{eqnarray}
\begin{pmatrix}
 v^2\\
 v_s^2
\end{pmatrix}
=
 \frac{2}{ 4 \lambda_{1\Phi} \lambda_{s1} - \lambda_{s\Phi 1}^2 }
 \begin{pmatrix}
  2 \lambda_{s1} & -\lambda_{s\Phi 1}\\
  -\lambda_{s\Phi 1} & 2 \lambda_{1\Phi}
 \end{pmatrix}
 \begin{pmatrix}
  m_\Phi^2\\
  m_{s_1}^2
 \end{pmatrix} .
\end{eqnarray}
 The $Z_2$-odd scalars in this model are
two CP-even neutral ones
(${\mathcal H}_1^0$ and ${\mathcal H}_2^0$),
a CP-odd neutral one (${\mathcal A}^0 = \eta_i^0$),
and a charged pair (${\mathcal H}^\pm = \eta^\pm$).
The CP-even scalars are defined as
\begin{eqnarray}
\begin{pmatrix}
 {\mathcal H}_1^0\\
 {\mathcal H}_2^0
\end{pmatrix}
=
\begin{pmatrix}
 \cos\theta_0^\prime & -\sin\theta_0^\prime\\
 \sin\theta_0^\prime & \cos\theta_0^\prime
\end{pmatrix}
\begin{pmatrix}
 \eta_r^0\\
 s_2^0
\end{pmatrix} , \quad
\tan{2\theta_0^\prime}
=
 \frac{ \sqrt{2}\, \lambda_{s\Phi\eta}\, v\, v_s }
      {
       ({\mathcal M}_0)_{ss}^2 - ({\mathcal M}_0)_{\eta\eta}^2
      } ,
\end{eqnarray}
where
$({\mathcal M}_0)_{\eta\eta}^2 \equiv
m_\eta^2
+ ( \lambda_{1\Phi\Phi} + \lambda_{1\Phi\eta} )\, v^2/2
+ \lambda_{s\eta 1}\, v_s^2/2$
and
$({\mathcal M}_0)_{ss}^2 \equiv
m_{s_2^0}^2
+ \lambda_{s3}\, v_s^2
+ \lambda_{s\Phi 2}\, v^2$.
 Squared masses of these scalars are given by
\begin{eqnarray}
m_{{\mathcal H}_1^0}^2
&=&
 \frac{1}{\,2\,}
 \left\{
  ({\mathcal M}_0)_{\eta\eta}^2
  + ({\mathcal M}_0)_{ss}^2
  - \sqrt{
          \bigl\{
           ({\mathcal M}_0)_{\eta\eta}^2 - ({\mathcal M}_0)_{ss}^2
          \bigr\}^2
          + 2\, \lambda_{s\Phi\eta}^2\, v^2\, v_s^2 \
         }
 \right\} ,
\label{eq:mH1}\\
%
%
m_{{\mathcal H}_2^0}^2
&=&
 \frac{1}{\,2\,}
 \left\{
  ({\mathcal M}_0)_{\eta\eta}^2
  + ({\mathcal M}_0)_{ss}^2
  + \sqrt{
          \bigl\{
           ({\mathcal M}_0)_{\eta\eta}^2 - ({\mathcal M}_0)_{ss}^2
          \bigr\}^2
          + 2\, \lambda_{s\Phi\eta}^2\, v^2\, v_s^2 \
         }
 \right\} ,
\label{eq:mH2}\\
m_{{\mathcal A}^0}^2
&=&
 ({\mathcal M}_0)_{\eta\eta}^2 , 
\label{eq:mA}\\
m_{{\mathcal H}^\pm}^2
&=&
 ({\mathcal M}_0)_{\eta\eta}^2
 - \frac{1}{\,2\,} \lambda_{1\Phi\eta}\, v^2 .
\label{eq:mHpm}
\end{eqnarray}
 Notice that
$m_{{\mathcal H}_1^0} \leq m_{{\mathcal A}^0} \leq m_{{\mathcal H}_2^0}$.
 We assume $m_{{\mathcal H}_1^0} < m_{{\mathcal H}^\pm}$
and then
${\mathcal H}_1^0$ becomes the dark matter candidate.
 Hereafter
it is assumed that the mixing $\theta_0^\prime$ is small.

\begin{figure}[t]
\begin{center}
\includegraphics[scale=0.6]{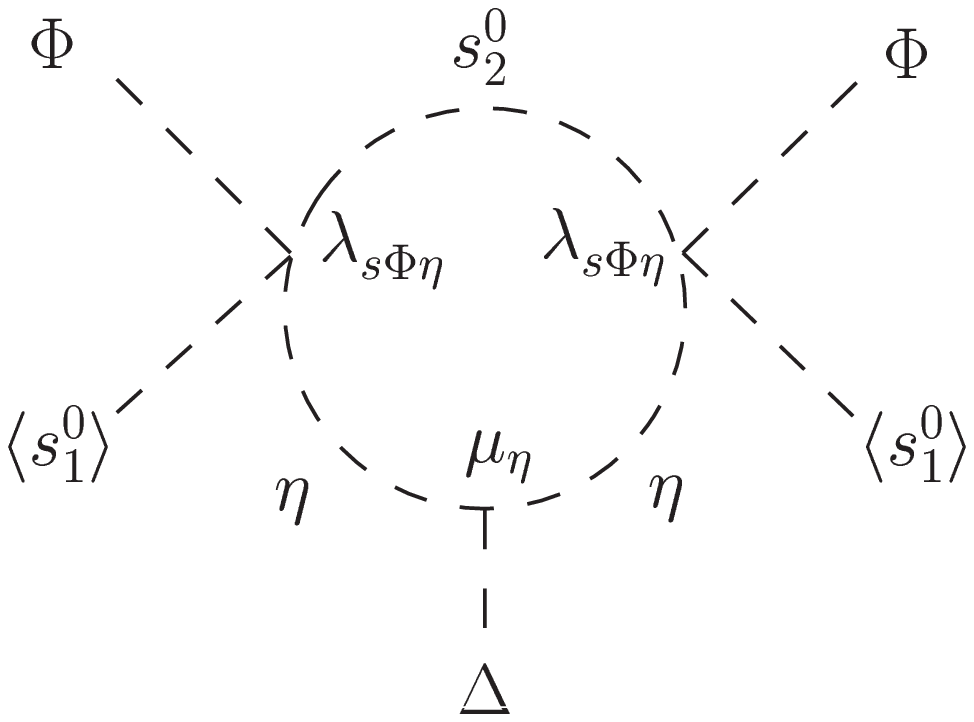}
\includegraphics[scale=0.3]{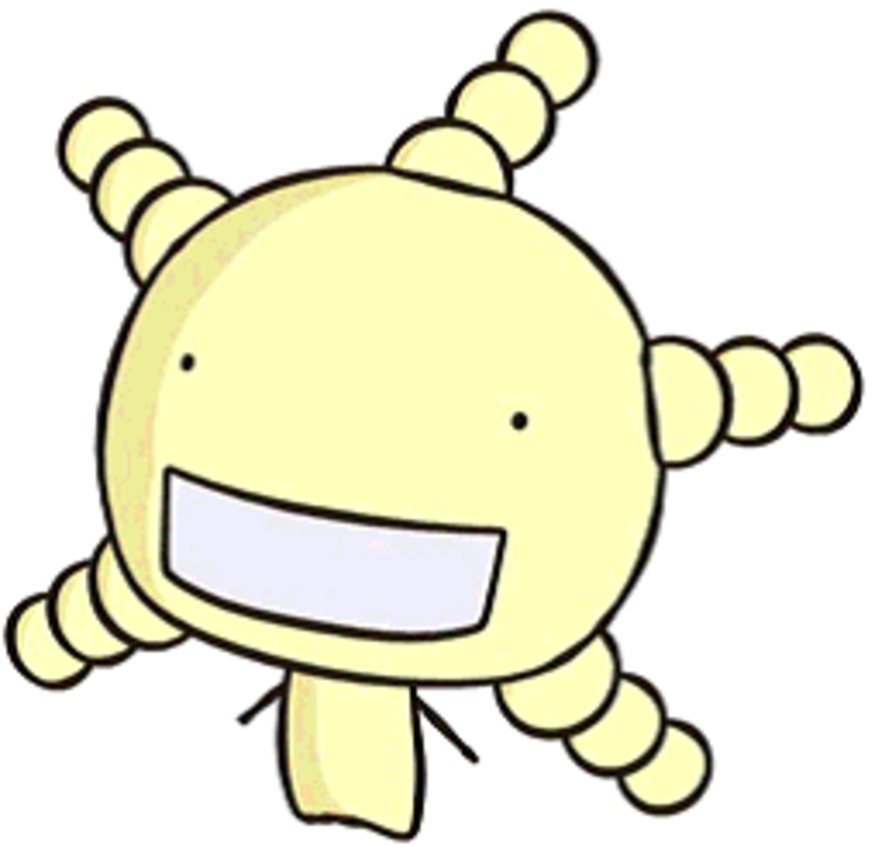}
\vspace*{-5mm}
\caption{
 One-loop diagram for the $\mu$-term.
 We call it "A.~oryzae diagram"~\cite{moyasimon}.
}
\label{fig:1-loop}
\end{center}
\end{figure}

 The $\mu$-term is generated
by the one-loop diagram.
 Figure~\ref{fig:1-loop} is the dominant one
in the case of small $\theta_0^\prime$.
 Then,
the parameter $\mu$ is calculated as
\begin{eqnarray}
\mu
&=&
 \frac{ \lambda_{s\Phi\eta}^2\, \mu_\eta^{} v_s^2 }
      { 64\pi^2
        \bigl\{
         ({\mathcal M}_0)_{ss}^2 - ({\mathcal M}_0)_{\eta\eta}^2
        \bigr\}
      }
 \left\{
  1
  - \frac{ ({\mathcal M}_0)_{ss}^2 }
         { ({\mathcal M}_0)_{ss}^2 - ({\mathcal M}_0)_{\eta\eta}^2 }
    \ln\frac{({\mathcal M}_0)_{ss}^2}{({\mathcal M}_0)_{\eta\eta}^2}
 \right\} .
\label{eq:mu-loop}
\end{eqnarray}
 The one-loop induced $\mu$ parameter
can be expected to be much smaller than $\mu_\eta$.
 The suppression factor $|\mu/\mu_\eta|$
is estimated in Sec.~\ref{subsec:DM}.

\section{phenomenology}
\label{sec:pheno}

\subsection{Dark matter}
\label{subsec:DM}

 If $({\mathcal M}_0)_{\eta\eta} < ({\mathcal M}_0)_{ss}$,
the dark matter candidate ${\mathcal H}_1^0$
is given by $\eta_r^0$ approximately
because we assume small mixing.
 See, e.g., Ref.~\cite{i-doublet}
for studies about the inert doublet scalar.
 Let us assume
$m_{{\mathcal H}_1^0}^{}\simeq 75\,\GeV$
and $m_{{\mathcal A}^0}^{} \gtrsim 125\,\GeV$.
 As shown in Ref.~\cite{Lundstrom:2008ai},
these values satisfy
constraints from the LEP experiments~\cite{:2005ema,EspiritoSanto:2003by}
and the WMAP experiment~\cite{Komatsu:2008hk}.
 The mass splitting
($m_{{\mathcal A}^0}^{} - m_{{\mathcal H}_1^0}^{} \gtrsim 50\,\GeV$)
suppresses
quasi-elastic scattering on nuclei
(${\mathcal H}_1^0 N \to {\mathcal A}^0 N$
mediated by the $Z$ boson)
enough to satisfy constraints
from direct search experiments of the DM~\cite{Aprile:2011hi}.
 By using eqs.~\eqref{eq:mH1} and \eqref{eq:mA},
we obtain
\begin{eqnarray}
\frac{ \lambda_{s\Phi\eta}^2 v_s^2 }{ ({\mathcal M}_0)_{ss}^2 }
\simeq
 \frac{ 2}{ v^2 }
 \left( m_{{\mathcal A}^0}^2 - m_{{\mathcal H}_1^0}^2 \right)
\gtrsim 0.3 .
\label{eq:lambda}
\end{eqnarray}
 In order to be consistent
with our assumption of small $\theta_0^\prime$
(e.g., $\simeq 0.1$),
$({\mathcal M}_0)_{ss} \gtrsim 3\,\TeV$ is required.
 The value in eq.~\eqref{eq:lambda} results in
\begin{eqnarray}
\frac{\mu}{\mu_\eta} \gtrsim 10^{-4} .
\label{eq:mu-50GeV}
\end{eqnarray}
 For the greater value of $m_{{\mathcal A}^0}^{}$,
the larger $\mu/\mu_\eta$ is predicted.
 In particular,
by taking $m_{{\mathcal A}^0}^{}$ to be the TeV scale,
we obtain $\mu/\mu_\eta \sim 10^{-2}$,
which yields $v_\Delta^{} \sim 1\,\GeV$
for $\mu_\eta^{}$ and $m_\Delta^{}$ to be
at the electroweak scale.
 Such a value for $v_\Delta^{}$
is suggested in the recent study of
radiative corrections to the electroweak parameters%
~\cite{Kanemura:2012rs}.

 On the contrary,
if we take $m_{{\mathcal A}^0}^{} \simeq 83\,\GeV$
which is allowed in a tiny region~\cite{Lundstrom:2008ai},
values in eqs.~\eqref{eq:lambda} and \eqref{eq:mu-50GeV}
become 10 times smaller.
 We mention that
the WMAP constraint might be changed
by a characteristic annihilation process
${\mathcal H}_1^0 {\mathcal H}_1^0
\to \Delta_r^0 \to \overline{\nu}\,\overline{\nu}$
where
${\mathcal H}_1^0 {\mathcal H}_1^0 (\Delta_r^0)^\ast$ interaction
is governed by $\mu_\eta^{}$
(not by a tiny $\mu$).
 This additional process could sift
allowed value of $m_{{\mathcal H}_1^0}^{}$ to lower one
while $m_{{\mathcal A}^0}^{} \gtrsim 100\,\GeV$
due to the LEP constraint.
 Then,
$\mu/\mu_\eta$ might become larger than
the value in eq.~\eqref{eq:mu-50GeV}
because of larger $m_{{\mathcal A}^0}^{} - m_{{\mathcal H}_1^0}^{}$.
 This undesired effect would be easily avoided
if $m_{\Delta_r^0}^{}$ is away enough
from $2 m_{{\mathcal H}_1^0}^{}$.

 On the other hand,
${\mathcal H}_1^0$
comes dominantly from $s_2^0$
if $({\mathcal M}_0)_{\eta\eta} > ({\mathcal M}_0)_{ss}$.
 See, e.g., Ref.~\cite{r-i-singlet}
for studies about the real inert singlet scalar.
 Coupling $\sqrt{2}\, \lambda_{s\Phi 1}\, v$
of the ${\mathcal H}_1^0 {\mathcal H}_1^0 h^0$ interaction
($h^0$ is the SM Higgs boson)
determines annihilation cross section of ${\mathcal H}_1^0$
and scattering cross section on nuclei.
 If we introduce the $\text{U}(1)_\BL$ gauge symmetry,
the scattering of $s_2^0$ on nuclei
can be mediated also by the gauge boson $Z^\prime$.
 Notice that
the parameter $\lambda_{s\Phi 1}$
(and also the $\text{U}(1)_\BL$ gauge coupling constant)
does not affect on $\mu$ parameter in eq.~\eqref{eq:mu-loop}.
 Let us estimate
the magnitude of $\mu/\mu_\eta$.
 In the usual HTM,
$h_{\ell\pell}$ is expected to be
$\lesssim 10^{-2}$ for $m_{H^{\pm\pm}} \sim 100\,\GeV$
in order to suppress LFV processes.
 Thus,
we may accept $\lambda_{s\Phi\eta} \sim 1 \text{-} 10^{-2}$
as a value which is not too small.
 Assuming
$({\mathcal M}_0)_{ss}
 \ll ({\mathcal M}_0)_{\eta\eta} \sim v_s \sim 1\,\TeV$
for example%
\footnote
{
 If we introduce $\text{U}(1)_\BL$ gauge symmetry
in order to eliminate the Majoron,
$v_s$ should be a little bit larger
(e.g., $\geq 3\,\TeV$)
due to constraint on the mass of $Z^\prime$.
},
we have a suppression factor as
\begin{eqnarray}
\left| \frac{\mu}{\mu_\eta} \right|
\sim
 \frac{ \lambda_{s\Phi\eta}^2 v_s^2 }
      { 64 \pi^2 ({\mathcal M}_0)_{\eta\eta}^2 }
\sim
 10^{-3} \text{-} 10^{-7} .
\end{eqnarray}
 Thus,
even if the value of $\mu_\eta$ is in the TeV scale,
we can obtain $\mu \sim 0.1\,\MeV$
although we need further suppression
with $h_{\ell\pell} \lesssim 10^{-5}$
to have $m_\nu \lesssim 1\,\eV$.
 If we use $h_{\ell\pell} \sim \lambda_{s\Phi\eta} \sim 10^{-3}$,
we obtain $|\mu/\mu_\eta| h_{\ell\pell} \sim 10^{-12}$
which can connect the TeV scale $\mu_\eta$
to the eV scale $m_\nu$.

\subsection{Collider}
\label{subsec:col}

\begin{figure}[t]
\begin{center}
\includegraphics[scale=0.8]{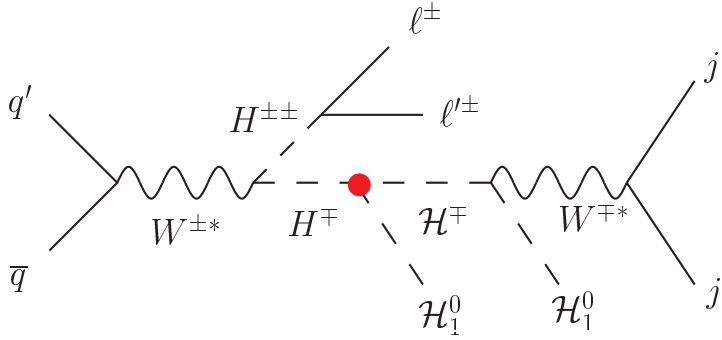}
\vspace*{-5mm}
\caption{
 The unique process in our model for ${\mathcal H}_1^0 \simeq \eta^0_r$.
 The bosonic decay of $H^{+}$
contains information of $\mu_\eta$
indicated by a red blob.
}
\label{fig:LHC}
\end{center}
\end{figure}

 The characteristic feature of our model is that
$\mu_\eta$ is much larger than $\mu$.
 Let us consider possibility to
probe the large $\mu_\eta$ in collider experiments.

 A favorable process is shown in Fig.~\ref{fig:LHC}
for ${\mathcal H}_1^0 \simeq \eta^0_r$.
 For simplicity,
we take $\lambda_5=0$
which results in
$m_{H^{\pm\pm}}^{} \simeq m_{H^\pm}^{}
\simeq m_{H^0,A^0}^{}$.
 Recently,
it was found in Ref.~\cite{Kanemura:2012rs}
that the electroweak precision test prefers $\lambda_5 > 0$
in the HTM where the electroweak sector
is described by four input parameters.
 However,
results in Ref.~\cite{Kanemura:2012rs}
might not be applied directly to our model%
\footnote
{
 Our model also has four parameters
for the electroweak sector
although $v_\Delta$ is generated at the 1-loop level. 
}
because the scalar sector is extended.
 Since 
$H^{\pm\pm} \to \ell^\pm {\ell^\prime}^\pm$
is the most interesting decay in the HTM,
we assume $2 m_{{\mathcal H}^\pm}^{} > m_{H^{\pm\pm}}^{}$
in order to forbid $H^{\pm\pm} \to {\mathcal H}^\pm {\mathcal H}^\pm$.
 Even in this case,
the DM ${\mathcal H}_1^0$
can be light enough
($m_{H^\pm}^{} > m_{{\mathcal H}^\pm}^{} + m_{{\mathcal H}_1^0}^{}$)
so that $Z_2$-even charged scalar $H^\pm$ ($\simeq \Delta^\pm$)
can decay into ${\mathcal H}^\pm {\mathcal H}_1^0$
via $\mu_\eta$-term
which is indicated by a red blob in Fig.~\ref{fig:LHC}.
 The partial decay width of
$H^\pm \to {\mathcal H}^\pm {\mathcal H}_1^0$
is determined by $(\mu_\eta/\mu)^2 v_\Delta^2/m_{H^\pm}^{}$
while the width of $H^\pm \to \ell^\pm \nu$
is proportional to $m_{H^\pm}^{} m_\nu^2/v_\Delta^2$.
 Taking $\mu_\eta/\mu \sim 10^4$,
$v_\Delta \sim 10\,\keV$,
$m_{H^\pm}^{} \sim 100\,\GeV$,
and $m_\nu^{} \sim 0.1\,\eV$
for example,
we have $(\mu_\eta/\mu)^2 v_\Delta^2/m_{H^\pm}^{} \sim 10^5\,\eV$
and $m_{H^\pm}^{} m_\nu^2/v_\Delta^2 \sim 10\,\eV$.
 Then,
$H^\pm$ dominantly decays into ${\mathcal H}^\pm {\mathcal H}_1^0$.
 Finally,
${\mathcal H}^\pm$ decays into $(W^\pm)^\ast {\mathcal H}_1^0$.
 Therefore,
from a production mechanism
$pp \to (W^\pm)^\ast \to H^{\pm\pm} H^\mp$,
we would have
$\ell\ell j j \cancel{E}_T$
as a final state%
\footnote
{
 Each of two ${\mathcal H}_1^0$ in Fig.~\ref{fig:LHC}
can be replaced with ${\mathcal A}^0$
which decays into $Z^* {\mathcal H}_1^0$
for ${\mathcal H}_1^0 \simeq \eta^0_r$.
}
for which $\ell\ell$ has the invariant mass $m(\ell\ell)$
at $m_{H^{\pm\pm}}^{}$
assuming that the value of $m_{H^{\pm\pm}}^{}$
has been known already.

 If ${\mathcal H}_1^0 \simeq s_2^0$,
then $H^\pm$ decays via $\mu_\eta$-term
into ${\mathcal H}^\pm {\mathcal A}^0$
or ${\mathcal H}^\pm {\mathcal H}_2^0$
followed by ${\mathcal H}_2^0 \to {\mathcal A}^0 J^0$
where a sizable $\lambda_{s\eta 1}$ is assumed%
\footnote
{
 If $\lambda_{s\eta 1}$ is small,
${\mathcal H}_2^0$~($\simeq \eta_r^0$)
decays into $Z^\ast {\mathcal A}^0$.
}.
 Because of ${\mathcal A}^0 \to {\mathcal H}_1^0 J^0$
through $\lambda_{s\Phi\eta}$,
we have again
$\ell\ell j j \cancel{E}_T$ with $m(\ell\ell) = m_{H^{\pm\pm}}^{}$
from $pp \to (W^\pm)^\ast \to H^{\pm\pm} H^\mp$.

 In the usual HTM in contrast,
the final state with such $\ell\ell$
is likely to include additional charged leptons
($\ell\ell\ell\ell$ from $H^{++} H^{--}$,
$\ell\ell\ell\cancel{E}_T$ from $H^{\pm\pm} H^\mp$,
etc.)
if $H^{\pm\pm}$ decay dominantly into $\ell^\pm\ell^{\prime\pm}$.
 Therefore,
our model would be supported
if experiments observe
final states which include jets and only two $\ell$
whose invariant mass gives $m(\ell\ell) = m_{H^{\pm\pm}}^{}$.
 This potential signature
might be disturbed by hadronic decays of $\tau$
because $H^{++}H^{--} \to \ell\ell\tau\tau$ can result in
$\ell\ell j j \cancel{E}_T$ with $m(\ell\ell) = m_{H^{\pm\pm}}^{}$.
 Realistic simulation
is necessary to see the feasibility.

\section{conclusions and discussion}
\label{sec:concl}

 We have presented
the simple extension of the HTM
by introducing
a $Z_2$-even neutral scalar $s_1^0$ of $L\#=-1$,
a $Z_2$-odd neutral real scalar $s_2^0$ of $L\#=0$,
and a $Z_2$-odd doublet scalar field $\eta$ of $L\#=-1$.
 The DM candidate ${\mathcal H}^0_1$ in our model
is made from $s_2^0$ and $\eta_r$.
 The $\mu \Phi^T i\sigma_2 \Delta^\dagger \Phi$ interaction
which is the origin of $v_\Delta$ (and neutrino masses)
is induced at the one-loop level
while
the $\mu_\eta^{} \eta^T i\sigma_2 \Delta^\dagger \eta$ interaction
exists at the tree level.
 Because of the loop suppression for $\mu$ parameter,
the model gives small neutrino masses naturally
without using very heavy particles.

 For ${\mathcal H}^0_1 \simeq \eta^0_r$,
the suppression factor $|\mu/\mu_\eta^{}|$
is constrained by the DM relic abundance measured
by the WMAP experiment.
 We have shown that
$|\mu/\mu_\eta^{}| \sim 10^{-4}\,\text{-}\,10^{-5}$
is possible.
 On the other hand,
for ${\mathcal H}^0_1 \simeq s_2^0$,
the suppression factor is somewhat free from
experimental constraints on the DM\@.
 In our estimate,
$|\mu/\mu_\eta^{}| \sim 10^{-3} \text{-} 10^{-7}$
can be obtained as an example
with $\lambda_{s\Phi\eta} \sim 1 \text{-} 10^{-2}$.

 The characteristic feature of the model
is that $\mu_\eta$ is not small
while $\mu$ can be small.
 A possible collider signature
which depends on $\mu_\eta$
would be $\ell\ell j j \cancel{E}_T$
with the invariant mass $m(\ell\ell) = m_{H^{\pm\pm}}^{}$
because more charged leptons are likely
to exist in such final states
in the usual HTM\@.

\begin{acknowledgments}
 The work of S.K.\ was supported by
Grant-in-Aid for Scientific Research 
Nos.\ 22244031 and 23104006.
 The work of H.S.\ was supported
by the Sasakawa Scientific Research Grant
from the Japan Science Society
and Grant-in-Aid for Young Scientists (B)
No.\ 23740210.
\end{acknowledgments}

\appendix

\section*{appendix}

 The Higgs potential of our model is
given by $V=V_2+V_3+V_4$ where
\begin{eqnarray}
V_2
&\equiv&
 - m_{s_1}^2 |s_1^0|^2
 + \frac{1}{\,2\,} m_{s_2}^2 (s_2^0)^2
 - m_\Phi^2\, \Phi^\dagger \Phi
 + m_\eta^2\, \eta^\dagger \eta 
 + m_\Delta^2\, \tr(\Delta^\dagger \Delta) ,
\end{eqnarray}
%
\begin{eqnarray}
V_3
&\equiv&
 ( \mu_\eta^{}\, \eta^T\, i\sigma_2\, \Delta^\dagger\, \eta)
 + \text{h.c.} ,
\end{eqnarray}
%
\begin{eqnarray}
V_4
&\equiv&
 \lambda_{1\Phi}\, (\Phi^\dagger \Phi)^2
 + \lambda_{1\eta}\, (\eta^\dagger \eta)^2
 + \lambda_{1\Phi\Phi}\, (\Phi^\dagger \Phi) (\eta^\dagger \eta)
 + \lambda_{1\Phi\eta}\, (\Phi^\dagger \eta) (\eta^\dagger \Phi)
\nonumber\\
&&{}
 + \lambda_2\, [\tr(\Delta^\dagger \Delta)]^2
 + \lambda_3\, \tr[(\Delta^\dagger \Delta)^2]
\nonumber\\
&&{}
 + \lambda_{4\Phi}\,
   (\Phi^\dagger \Phi)\, \tr(\Delta^\dagger \Delta)
 + \lambda_{4\eta}\,
   (\eta^\dagger \eta)\, \tr(\Delta^\dagger \Delta)
\nonumber\\
&&{}
 + \lambda_{5\Phi}\,
   (\Phi^\dagger \Delta \Delta^\dagger \Phi)
 + \lambda_{5\eta}\,
   (\eta^\dagger \Delta \Delta^\dagger \eta)
\nonumber\\
&&{}
 + \lambda_{s1}\, |s_1^0|^4
 + \lambda_{s2}\, (s_2^0)^4
 + \lambda_{s3}\, |s_1^0|^2 (s_2^0)^2
\nonumber\\
&&{}
 + \lambda_{s\Phi 1}\, |s_1^0|^2\, (\Phi^\dagger \Phi)
 + \lambda_{s\Phi 2}\, (s_2^0)^2\, (\Phi^\dagger \Phi)
\nonumber\\
&&{}
 + \lambda_{s\eta 1}\, |s_1^0|^2\, (\eta^\dagger \eta)
 + \lambda_{s\eta 2}\, (s_2^0)^2\, (\eta^\dagger \eta)
 + \left\{
    \lambda_{s\Phi\eta}\, s_1^0\, s_2^0\, (\eta^\dagger \Phi)
    + \text{h.c.}
   \right\}
\nonumber\\
&&{}
 + \lambda_{s\Delta 1}\, |s_1^0|^2 \tr(\Delta^\dagger \Delta)
 + \lambda_{s\Delta 2}\, (s_2^0)^2 \tr(\Delta^\dagger \Delta) .
\end{eqnarray}
 All coupling constants are real
because the phases of $\mu_\eta$ and $\lambda_{s\Phi\eta}$
can be absorbed by $\Delta$ and $s_1^0$, respectively.

 Mass eigenstates of two $Z_2$-even CP-even neutral scalars
which are composed of $s_{1r}^0$ and $\phi_r^0$
are obtained as
\begin{eqnarray}
\begin{pmatrix}
h^0\\
H^0
\end{pmatrix}
=
 \begin{pmatrix}
  \cos\theta_0 & -\sin\theta_0\\
  \sin\theta_0 & \cos\theta_0
 \end{pmatrix}
 \begin{pmatrix}
  \phi^0_r\\
  s_{1r}^0
 \end{pmatrix} , \quad
%
\tan{2\theta_0}
=
 \frac{ \lambda_{s\Phi 1}\, v\, v_s }
      { \lambda_{s1} v_s^2 - \lambda_{1\Phi} v^2 } .
\end{eqnarray}
 Their masses eigenvalues are given by
\begin{eqnarray}
m_{h^0}^2
&\simeq&
 \lambda_{1\Phi} v^2 + \lambda_{s1} v_s^2
 - \sqrt{
         \left( \lambda_{1\Phi} v^2 - \lambda_{s1} v_s^2 \right)^2
         + \lambda_{s\Phi 1}^2 v^2\, v_s^2
        } \ ,\\
%
%
m_{H^0}^2
&\simeq&
 \lambda_{1\Phi} v^2 + \lambda_{s1} v_s^2
 + \sqrt{
         \left( \lambda_{1\Phi} v^2 - \lambda_{s1} v_s^2 \right)^2
         + \lambda_{s\Phi 1}^2 v^2\, v_s^2
        } \ ,
\end{eqnarray}
where small contributions from $v_\Delta^{}$ are neglected. 
 Two $Z_2$-even CP-odd neutral bosons
($\phi_i^0$ and $s_{1i}^0$) are Nambu-Goldstone bosons;
 $\phi_i^0$ is absorbed by the $Z$ boson,
and $s_{1i}^0$ is the Majoron (or absorbed by the $Z^\prime$ boson).

 Masses of bosons made dominantly from $\Delta$ are given by
\begin{eqnarray}
m_{H_T^0}^2 \simeq m_{A_T^0}^2
&\simeq&
 m_{H^\pm}^2
 + \frac{1}{\,4\,} \lambda_{5\Phi} v^2 ,\\
%
%
m_{H^\pm}^2
&\simeq&
 m_\Delta^2
 + \frac{1}{\,4\,}
   ( 2\lambda_{4\Phi} + \lambda_{5\Phi} ) v^2
 + \frac{1}{\,2\,} \lambda_{s\Delta 1} v_s^2 ,\\
%
%
m_{H^{\pm\pm}}^2
&\simeq&
 m_{H^\pm}^2
 - \frac{1}{\,4\,} \lambda_{5\Phi} v^2 .
\end{eqnarray}


\end{document}